# Measuring the Knowledge-Based Economy of China in terms of Synergy among Technological, Organizational, and Geographic Attributes of Firms

Loet Leydesdorff [a] & Ping Zhou [b]*

**Abstract**

Using the possible synergy among geographic, size, and technological distributions of firms in the *Orbis* database, we find the greatest reduction of uncertainty at the level of the 31 provinces of China, and an additional 18.0% at the national level. Some of the coastal provinces stand out as expected, but the metropolitan areas of Beijing and Shanghai are (with Tianjin and Chongqing) most pronounced at the next-lower administrative level of (339) prefectures, since these four "municipalities" are administratively defined at both levels. Focusing on high- and medium-tech manufacturing, a shift toward Beijing, Shanghai, and Tianjin (near Beijing) is indicated, but the synergy is on average not enhanced. High- and medium-tech manufacturing is less embedded in China than in Western Europe. Knowledge-intensive services "uncouple" the knowledge base from the regional economies mostly in Chongqing and Beijing. Unfortunately, the *Orbis* data is incomplete since it was collected for commercial and not for administrative or governmental purposes. However, we provide a methodology that can be used by others who may have access to higher-quality statistical data for the measurement.

---

[a] Amsterdam School of Communication Research (ASCoR), University of Amsterdam, Kloveniersburgwal 48, 1012 CX Amsterdam, The Netherlands; loet@leydesdorff.net ; http://www.leydesdorff.net
[b] Department of Information Resource Management, Zhejiang University, No. 866 Yuhangtang Road, Hangzhou, 310058, China; pingzhou@zju.edu.cn ; * corresponding author

# 1. Introduction

Mutual information in three (or more) dimensions can be derived from Shannon's (1948) formulas of information theory (e.g., McGill, 1954; Abramson, 1963: 131 ff.), but it can no longer be considered as Shannon-type information because it is a signed information measure that can also be negative (Krippendorff, 2009a; Yeung, 2008: 59f.). Any two sources of variance in the three- (or more-)way interaction may spuriously correlate given a third variable because of mutual overlaps in the expected information contents of the distributions. The overlaps can be considered as repetitions and therefore redundant or, alternatively, these non-linearities may be considered as loops that are incompatible with the linear framework of Shannon's theory (Krippendorff, 2009b).

Leydesdorff & Ivanova (in press) have recently shown that the information in the overlaps should be counted twice—as in the case of pure sets—so that the total information content is enlarged, and consequently the complementary relation between information and redundancy is shifted in the favor of the redudancy. Unlike Krippendorff's (2009a) interaction information ($I_{ABC \rightarrow AB:AC:BC}$), mutual information in more than two dimensions can then be considered consistently as a measure of redundancy or reduction of the uncertainty that prevails at the systems level (Ivanova & Leydesdorff, in preparation).

Leydesdorff (2003) used this measure first as an operationalization of the possible reduction of uncertainty in the Triple Helix of university-industry-government relations (Park *et al.*, 2005; cf. Leydesdorff & Sun, 2009; Park & Leydesdorff, 2010; Ye & Leydesdorff, in press). When all relations are in place, an overlay may reduce uncertainty for the individual players at the systems level. This configurational effect cannot be attributed to specific links or nodes, but is a result of the interaction. These studies used co-authored publications as units of analysis, and considered possible relations as attributes of these units.

In another series of studies of European innovation systems we used three (or more) attributes of firms as a potential source of synergy in the economy. Storper (1997) hypothesized that the relational interaction among technology, geography, and organization can generate synergy in what he called a "Holy Trinity." We operationalize geography as the distribution in terms of geographical addresses, technology in terms of the OECD classification of firms according to the "Nomenclature générale des Activités économiques dans les Communautés Européennes" (NACE),[1] and organizational size in terms of numbers of employees. Can a latent construct among the three (or more) attributes be indicated that potentially reduces uncertainty in the configuration?

In the Netherlands (Leydesdorff *et al.*, 2006) and Sweden (Leydesdorff & Strand, in press), we found a regional pattern with surplus value at the national level, measurable as a between-regional reduction of uncertainty. In Germany (Leydesdorff & Fritsch, 2006) such national synergy was not found, but the surplus is realized at the level of the federal states, whereas the Hungarian system seems to consist of a western part integrated with neighboring EU-countries, a metropolitan center around Budapest, and an eastern part in which the old (state-led) system still prevails (Lengyel & Leydesdorff, 2011).

---

[1] The NACE code can be translated into the International Standard Industrial Classification (ISIC) that is used, for example, in the USA.

Most interestingly, Strand & Leydesdorff (2013) have found that the synergy in the Norwegian economy is concentrated along the shore in relation to marine and maritime (offshore) industries, whereas the university centers of the country (e.g., in Trondheim and Oslo) have not been integrated into the economy, but remain at a distance. In Sweden, 45.3% of the synergy was highly concentrated in the three metropolitan regions of Stockholm, Gothenburg, and Malmö/Lund.

Our design is based on using firms as units of analysis, and the following three variables and their interactions as attributes:

1. geographical addresses as an indicator of regional or other geographic provenance;
2. size in terms of number of employees as a proxy of economic organization (e.g., small and medium-sized companies);
3. NACE codes of the OECD as a technological classification.

In each study, however, we had to make compromises because of possible imperfections in the sample data. In Hungary, for example, we did not have NACE codes for all firms, and Statistics Sweden uses its own classification system. In these various studies, the data about numbers of employees was not always as fine-grained as it was in the original study of the Netherlands (Leydesdorff *et al.*, 2006).

In this further study we make an attempt to analyze the knowledge-based economy of China. Recently we obtained access to the database *Orbis* of the Bureau van Dijk (BvD; available at https://orbis.bvdinfo.com) containing the necessary data given our design. Although organized at the firm level and in considerable detail, this data is collected worldwide for commercial purposes and not for administrative or governmental purposes such as by a national bureau of statistics.

The data collection is based on more than one hundred information providers, or, as the BvD claims: "We're experts in company information and business intelligence. We integrate information from numerous sources to create *Orbis* and complement it with our own research. We combine this unique dataset with our software to create a dynamic global research tool."[2] The encompassing database covers company information for more than 100 million firms (including banks) worldwide. The three variables that are core to our research question (addresses, NACE codes, and numbers of employees) are also included.

*Orbis* provides information collected during the past ten years, but with a continuously moving window retrospectively from the current date. Regional databases are derived from *Orbis* (such as *Amadeus* for Europe), but contain approximately the same data only for the most recent year (and are therefore cheaper). The update frequencies of either database, however, are not specified precisely and may vary among countries and firms. In a study of the coverage of *Orbis*, Ribeiro *et al*. (2010) concluded that this coverage is poor, can vary among countries and sectors, but precise estimates were not provided. In any case, self-employed entrepreneurs are usually not covered, whereas these firms (e.g., startups) may be most interesting from the perspective of innovation policies. A further issue may be that headquarters and research centers are not always located in the same place, but we have the impression that *Orbis* tries to correct for such problems.

---

[2] "Orbis: Company Information around the Globe"; available at http://www.bvdinfo.com/About-BvD/Brochure-Library/Brochures/ORBIS-brochure.

In a recent study of the Italian innovation system (Cucco & Leydesdorff, in preparation) an almost perfect correlation was found (Pearson $r = 0.98$; Spearman's $\rho > 0.99$) between the distributions of our synergy values based on 462,316 valid observations using *Orbis* data versus 4,480,473 firms registered by Statistics Italy in 2007. This gives us some confidence in the representativeness of the *Orbis* data and its usefulness as a sample for our purpose of estimating synergy as a systems property. In this study, we explore the usefulness of this data as a sample for investigating the important question of how to measure the knowledge base of China in terms of Triple-Helix relations at the level of firm data. In any case, this study provides a methodology that can be improved if the complete data set for the population is made available.

We asked the National Bureau of Statistics of China for the complete set, but, for legal reasons, data is made available by this office only on the aggregated level at http://219.235.129.58/welcome.do#. Since no other data is readily available for China, we decided to explore *Orbis* as a source and thus harvested all firm data for China on December 14, 2012. With the caveat that this data is an incomplete sample because collected for business purposes as different from administrative purposes (such as governmental statistics), the domain enables us nonetheless to address in considerable detail the research question of where synergy is generated in the Chinese economy in terms of relations between geographical addresses, company sizes, and NACE codes.

## 2. Methods and materials

*2.1 Data*

Retrieval from *Orbis* provided us with 768,949 records with a Chinese address of which 768,948 could be downloaded. Figure 1 shows the yearly distribution of this data. (As noted, *Orbis* accumulates data for the last ten years.)

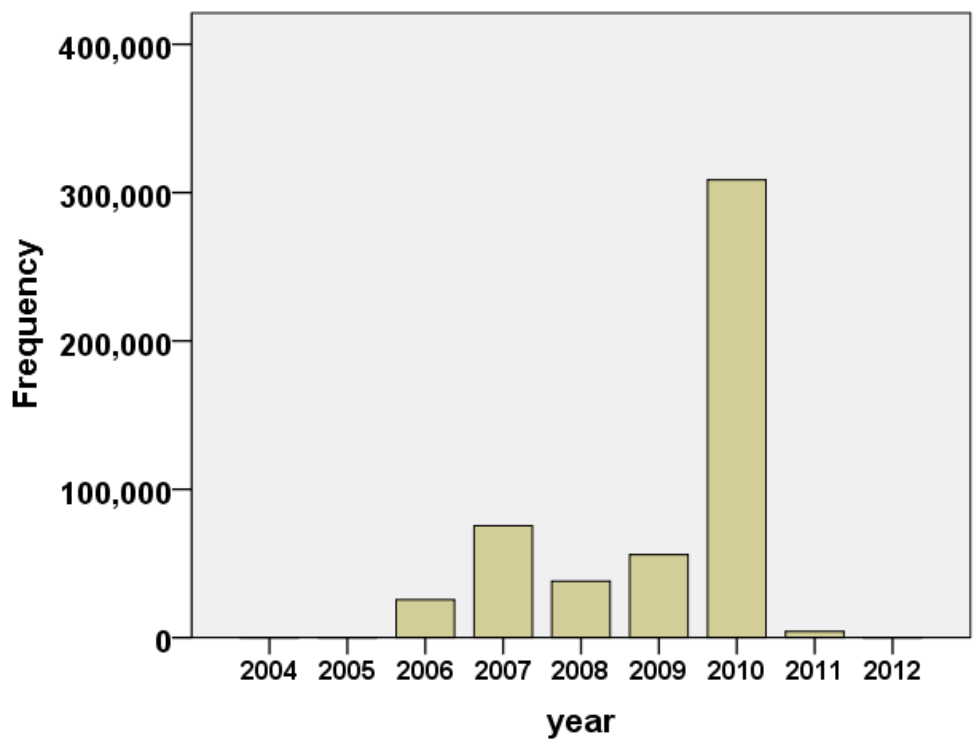

**Figure 1**: Yearly distribution of firm data for China from ORBIS (Dec. 14, 2012; *N* = 768,948).

Figure 1 teaches us that data collection in China did not begin until 2006, and that the data for 2011 were not yet included at the time of this download (Dec. 2012).[3] We focus on the data for 2008-2010, comprising 402,604 records (52.4%), of which 379,026 contained valid information in the three fields of interest: city name, NACE code (Rev. 2; 4 digits), and firm size.

---
[3] When we returned to the database on May 20, 2013, the retrieval was 1,612,309.

The attribution of city names to 31 provinces was fully standardized and made complete in terms of name variants, etc., by one of us. Name variants of cities were additionally standardized on the basis of computer routines that, for example, relabeled "Beijing Capital City" as equivalent to "Beijing", etc. As could be expected, the distribution of the firms included varies widely by region: from 67 for the province of Xizang (that is, the autonomous region of Tibet) to 62,805 for the heavily industrialized region of Jiangsu on the east coast (Figure 2). The distribution accords with the conventional wisdom that China is both industrialized and in large parts also rural.

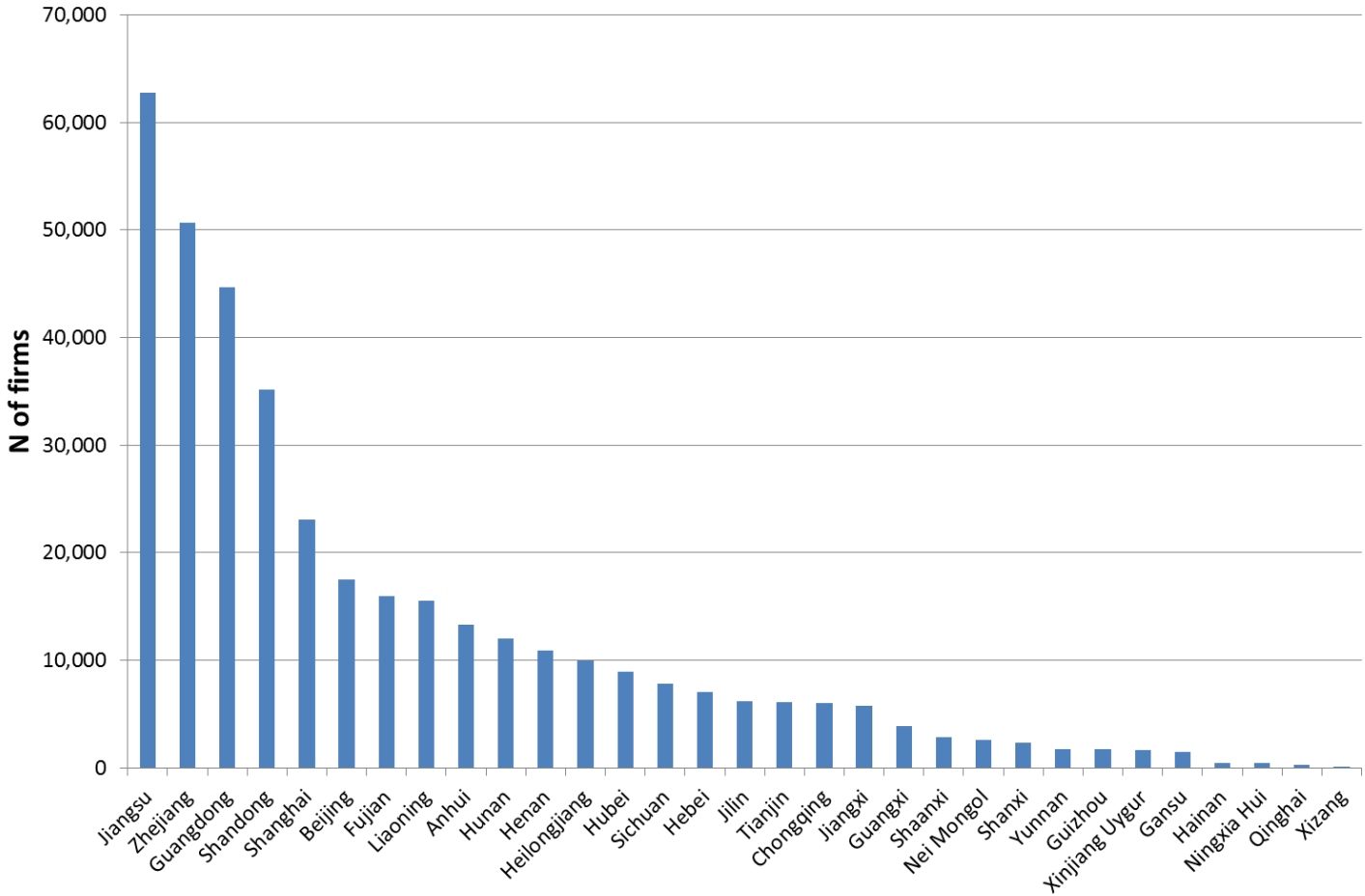

**Figure 2:** Geographical distribution of the firms (years 2008-2010; $N = 379,026$).

Using SPSS v21 for the geographic mapping, we imported a shapefile for the administrative organization of China from the Internet at http://www.diva-gis.org/datadown. The database of this file distinguishes three authoritative layers within China, of which the 31 provinces are the first, and 2,410 locations (city names) the third. The in-between level 2 contains 339 units (prefectures). Using dedicated routines, we were able to establish 330,897 records of firms (87.3%) additionally with this information at level 2.

For the size distribution in terms of numbers of employees, we used the same categories as those used in the first study of Dutch data (Leydesdorff *et al.*, 2006). Table 1 shows this distribution. As could be expected, medium-sized firms dominate the pattern; the relative absence of very small firms (0.9%) can be considered as an artifact of the data collection by the Bureau van Dijk (Ribeiro *et al.*, 2010). As noted, these missing values probably constitute the main shortcoming of this data because one would expect this class of firms to be relatively large.

| Size class | N | % |
|---|---|---|
| 0, 1, or n.a. | 3,761 | 0.9 |
| 2-4 | 1,003 | 0.2 |
| 5-9 | 4,027 | 1 |
| 10-19 | 19,938 | 5 |
| 20-49 | 95,949 | 23.8 |
| 50-99 | 110,662 | 27.5 |
| 100-199 | 83,469 | 20.7 |
| 200-499 | 58,271 | 14.5 |
| 500-749 | 10,676 | 2.7 |
| 750-999 | 4,870 | 1.2 |
| > 1000 | 9,978 | 2.5 |
| Total | 402604 | 100 |

**Table 1**: Size classes in terms of numbers of employees.

In 10,074 records (2.5%), the NACE-codes at the 2-digit level were not valid; these records were excluded from the analysis. *Orbis* uses Revision 2 of the NACE code at the four-digit level. Using the further classification of Eurostat/OECD (2009, 2011) in terms of high- and medium-tech manufacturing and knowledge-intensive services, 28,659 (or 7.1%) of the firms can be classified as high-tech manufacturing and 105,604 (26.2%) as medium-tech. We analyze this 33.3% of the records in a separate run of the data (Section 3.2). The NACE-categories of knowledge-intensive services, however, were populated only with 4,604 records (1.1%), and therefore less extensively analyzed.

*2.2. Methods*

As noted above, mutual information in more than three dimensions—the Triple-Helix indicator to be used here—is a signed information measure (Yeung, 2008), and therefore not a Shannon-information (Krippendorff, 2009a and b). However, this measure is derived in the context of information theory and follows from the Shannon formulas (e.g., Abramson, 1963; Ashby, 1964; McGill, 1954).

According to Shannon (1948) the uncertainty in the relative frequency distribution of a random variable $x$ ($\sum_x p_x$) can be defined as $H_X = -\sum_x p_x \log_2 p_x$. Shannon denotes this as probabilistic entropy, which is expressed in bits of information if the number two is used as the base for the logarithm. (When multiplied by the Boltzman constant $k_B$, one obtains thermodynamic entropy and the corresponding dimensionality in Joule/Kelvin. Unlike thermodynamic entropy, probabilistic entropy is dimensionless and therefore yet to be provided with meaning when a system of reference is specified.)

Likewise, uncertainty in a two-dimensional probability distribution can be defined as $H_{XY} = -\sum_x \sum_y p_{xy} \log_2 p_{xy}$. In the case of interaction between the two dimensions, the uncertainty is reduced with the mutual information or transmission: $T_{XY} = (H_X + H_Y) - H_{XY}$. If the distributions are completely independent $H_{XY} = H_X + H_Y$, and consequently $T_{XY} = 0$.

In the case of three potentially interacting dimensions (*x*, *y*, and *z*), the mutual information can be derived (e.g., Abramson, 1963: 131 ff.) as:

$$T_{XYZ} = H_X + H_Y + H_Z - H_{XY} - H_{XZ} - H_{YZ} + H_{XYZ} \tag{1}$$

The interpretation is as follows: association information can be categorized broadly into correlation information and interaction information. A spurious correlation in a third attribute, for example, can reduce the uncertainty between the other two. The correlation information among the attributes in a data set can be interpreted as the total amount of information shared among the attributes. The interaction information can be interpreted as multivariate dependencies among the attributes.

Compared with correlation, mutual information can be considered as a parsimonious measure for the association. The multivariate extension to mutual information was first introduced by McGill (1954) as a generalization of Shannon's mutual information. This *signed* information measure (Yeung, 2008: 59f.) is similar to the analysis of variance, but uncertainty analysis remains more abstract and does not require assumptions about the metric properties of the variables (Garner &

McGill, 1956). Han (1980) developed the concept further; positive and negative interactions were also discussed by Jakulin (2005), Leydesdorff & Ivanova (in press); Sun & Negishi (2010), Tsujishita (1995), and Yeung (2008: 59 f.).

One of the advantages of entropy statistics is that the values can be fully decomposed. As with the decomposition of probabilistic entropy (Theil, 1972: 20f.), mutual information in three dimensions can be decomposed into groups as follows:

$$T = T_0 + \sum_G \frac{n_G}{N} T_G \qquad (2)$$

Since we will decompose in the geographical dimension, $T_0$ connotes between-region uncertainty; $T_G$ the uncertainty prevailing at a geographical scale $G$; $n_G$ is the number of firms at this geographical scale; and $N$ the total number of firms in the whole set.

The between-group uncertainty ($T_0$) can be considered as a measure of the dividedness. A negative value of $T_0$ indicates an additional synergy at the higher level of national agglomeration among the lower-level geographical units. In the Netherlands, Norway, and Sweden, for example, a surplus was found at the national level; in Germany, this surplus was found at the level of the federal states (*Länder*). Note that one cannot compare the quantitative values of $T_0$ across countries—because these values are sample-specific—but one is allowed to compare the dividedness in terms of the positive or negative signs of $T_0$ and as a percentage of the total synergy for each country. All values of the contribution of subsets to the knowledge-based economy are based on normalization on the total set (that is, $n_G/N$ in Eq. 2).

## 3. Results

*3.1    Decomposition at the provincial level of China*

We run the analysis for all firms in the whole set and then for each of the 31 provinces separately. This leads to values for $T$ and $T_G$, respectively, that can be used in Eq. 2; the values of $N$ and $n_G$ are known from the download. Normalized values of the contributions of provinces to the national synergy ($\Delta T = {n_G}/{N} * T_G$) and the between-provinces synergy ($T_0$) can then be derived.

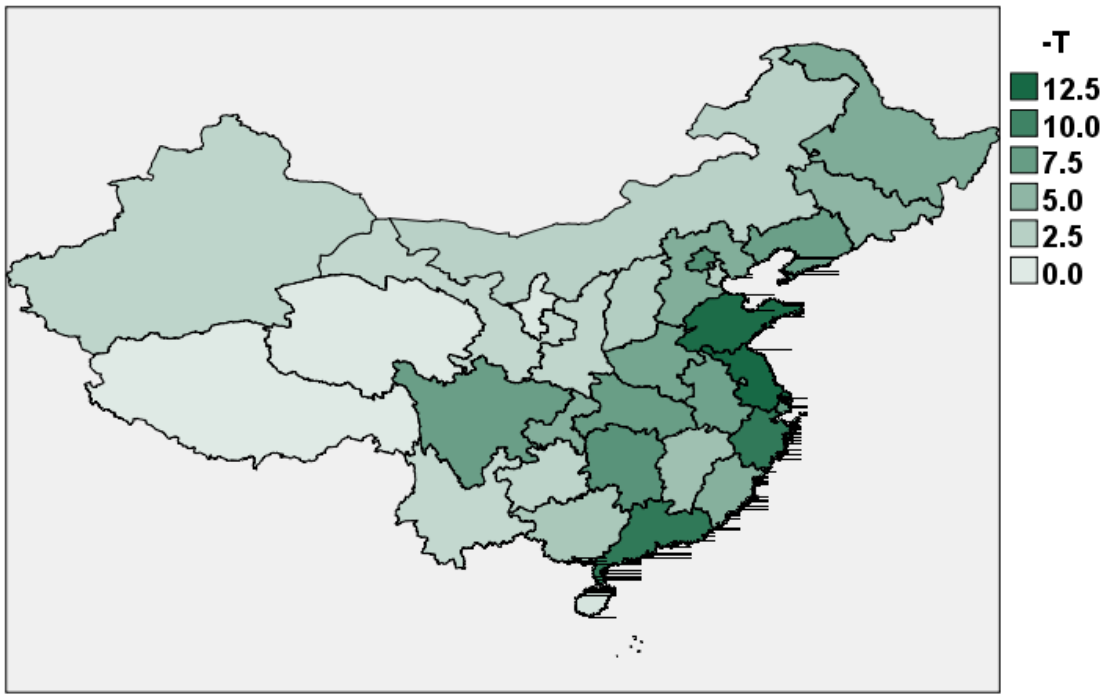

**Figure 3**: Synergies in the knowledge-based economy of China at the provincial level (years 2008-2010; $N = 379{,}026$).

Figure 3 provides a map of China with the 31 provincial units colored according to their respective contributions to the synergy in the knowledge-based economy. The total synergy for the nation is −196.48 mbits of information, of which 18.0% (35.46 mbits) is realized at the above-provincial level. This is more than we found in the case of Norway (11.7%), but less than for the Netherlands (27.1%) or Sweden (20.4%). As said, we found no additional synergy in Germany and Hungary at the national level (for different reasons). China thus functions very much as a unified nation state.

|  | All sectors | | High and medium tech | | | KIS | |
|---|---|---|---|---|---|---|---|
| Province (a) | N of firms (b) | ΔT (mbit) (c) | N of firms (d) | ΔT (mbit) (e) | %contr. [(e)/(c)] (f) | N of firms (g) | ΔT (mbit) (h) |
| Jiangsu | 62,805 | -12.48 | 1,783 | -3.19 | 25.6 | 49 | 0.03 |
| Shandong | 35,152 | -12.23 | 10,862 | -2.66 | 21.7 | 346 | 0.01 |
| Guangdong | 44,692 | -10.99 | 15,986 | -1.75 | 15.9 | 90 | -0.01 |
| Zhejiang | 50,699 | -10.92 | 17,265 | -2.33 | 21.4 | 24 | 0.05 |
| **Beijing** | 17,490 | -8.68 | 6,665 | -2.41 | **27.8** | 339 | **0.08** |
| Hunan | 12,019 | -8.38 | 3,811 | -1.80 | 21.5 | 218 | -0.12 |
| **Shanghai** | 23,049 | -7.93 | 10,164 | -2.37 | **29.9** | 17 | 0.01 |
| Hubei | 8,969 | -7.39 | 2,512 | -1.66 | 22.5 | 181 | -0.02 |
| Sichuan | 7,807 | -7.34 | 1,930 | -1.47 | 20.0 | 161 | -0.04 |
| Liaoning | 15,565 | -7.25 | 556 | -1.44 | 19.8 | 748 | -0.01 |
| Anhui | 13,275 | -6.98 | 4,260 | -1.61 | 23.0 | 130 | -0.01 |
| Henan | 10,899 | -6.62 | 3,088 | -1.46 | 22.0 | 112 | 0.01 |
| Heilongjiang | 9,993 | -5.99 | 3,174 | -1.22 | 20.4 | 137 | 0.04 |
| Hebei | 7,062 | -5.82 | 1,737 | -1.22 | 20.9 | 760 | 0.00 |
| **Chongqing** | 6,015 | -5.70 | 2,287 | -1.38 | 24.3 | 92 | **0.11** |
| Fujian | 16,001 | -5.59 | 3,344 | -1.17 | 21.0 | 187 | -0.03 |
| Jilin | 6,190 | -5.09 | 4,768 | -1.12 | 21.9 | 145 | 0.02 |
| Jiangxi | 5,790 | -4.01 | 1,977 | -0.98 | 24.3 | 140 | -0.02 |
| Guangxi | 3,888 | -3.33 | 1,112 | -0.65 | 19.7 | 89 | -0.07 |
| Shanxi | 2,363 | -2.72 | 698 | -0.72 | 26.7 | 75 | 0.01 |
| **Tianjin** | 6,132 | -2.69 | 2,774 | -0.83 | **30.8** | 13 | 0.00 |
| Nei Mongol | 2,605 | -2.37 | 24,823 | -0.28 | 11.6 | 147 | -0.01 |
| Guizhou | 1,695 | -2.11 | 433 | -0.23 | 10.8 | 31 | 0.03 |
| Xinjiang Uygur | 1,625 | -2.05 | 270 | -0.11 | 5.6 | 87 | 0.07 |
| Shaanxi | 2,868 | -1.92 | 971 | -0.45 | 23.4 | 96 | 0.02 |
| Gansu | 1,510 | -1.82 | 374 | -0.21 | 11.7 | 33 | 0.05 |
| Yunnan | 1,707 | -1.72 | 419 | -0.15 | 8.4 | 49 | 0.00 |
| Hainan | 431 | -0.54 | 111 | -0.05 | 9.9 | 12 | 0.01 |
| Ningxia Hui | 409 | -0.29 | 127 | -0.02 | 8.4 | 1 | 0.00 |
| Qinghai | 254 | -0.08 | 81 | 0.00 | - | 2 | 0.00 |
| Xizang (Tibet) | 67 | 0.01 | 12 | 0.00 | - | 6 | 0.01 |
| Σ = | | -161.02 | | -34.95 | 17.7 | | 0.23 |
| China | 379,026 | -196.48 | 128,374 | -41.75 | 21.2 | 4,517 | -1.07 |
| $T_0$ | | -35.46 | | -6.80 | 19.2 | | -1.30 |

**Table 1**: 31 Provinces of China sorted by their contribution to synergy (column c); all sectors (columns b-c; $N = 379{,}026$); high- and medium-tech manufacturing (column d-f; $n = 128{,}374$); knowledge-intensive services (columns g-h; $n = 4{,}517$). Total values of China and between-province values $T_0$ are added in the bottom rows. The four municipalities are boldfaced.

In Table 1, the 31 provinces are sorted in terms of their contributions to the overall synergy (column c). Since the number of firms in a region ($n_G$) is a factor in Eq. 2, the correlations with the number of firms are high, and therefore the relatively smaller provinces of Beijing and Shanghai (in terms of total numbers of firms) figure less prominently than one might perhaps expect. We return to this issue in Section 3.3 when we analyze the data at the next-lower level of aggregation. However, we first turn to the decomposition of the set in terms of high- and medium-tech manufacturing (columns d and e in Table 1), and knowledge intensive services (columns f and g) in the next section by using the NACE codes (Table 2).

**Table 2**: NACE classifications (Rev. 2) of high- and medium-tech manufacturing, and knowledge-intensive services.

| *High-tech Manufacturing* | *Knowledge-intensive Sectors (KIS)* |
|---|---|
| **21** Manufacture of basic pharmaceutical products and pharmaceutical preparations<br>**26** Manufacture of computer, electronic and optical products<br>**30.3** Manufacture of air and spacecraft and related machinery<br><br>*Medium-high-tech Manufacturing*<br><br>**20** Manufacture of chemicals and chemical products<br>**25.4** Manufacture of weapons and ammunition<br>**27** Manufacture of electrical equipment,<br>**28** Manufacture of machinery and equipment n.e.c.,<br>**29** Manufacture of motor vehicles, trailers and semi-trailers,<br>**30** Manufacture of other transport equipment<br>• **excluding 30.1** Building of ships and boats, and<br>• **excluding 30.3** Manufacture of air and spacecraft and related machinery<br>**32.5** Manufacture of medical and dental instruments and supplies | **50** Water transport,<br>**51** Air transport<br>**58** Publishing activities,<br>**59** Motion picture, video and television programme production, sound recording and music publishing activities,<br>**60** Programming and broadcasting activities,<br>**61** Telecommunications,<br>**62** Computer programming, consultancy and related activities,<br>**63** Information service activities<br>**64 to 66** Financial and insurance activities<br>**69** Legal and accounting activities,<br>**70** Activities of head offices; management consultancy activities,<br>**71** Architectural and engineering activities; technical testing and analysis,<br>**72** Scientific research and development,<br>**73** Advertising and market research,<br>**74** Other professional, scientific and technical activities,<br>**75** Veterinary activities<br>**78** Employment activities<br>**80** Security and investigation activities<br>**84** Public administration and defence, compulsory social security<br>**85** Education<br>**86 to 88** Human health and social work activities,<br>**90 to 93** Arts, entertainment and recreation<br><br>Of these sectors, **59 to 63, and 72** are considered *high-tech services*. |

Sources: Eurostat/OECD (2009, 2011); cf. Laafia (2002, p. 7) and Leydesdorff *et al.* (2006, p. 186).

*3.2 Sectorial decomposition*

*3.2.1 High- and medium-tech manufacturing*

Let us first turn to the subset of 134,263 firms (33.3% of the data) which are classified with the NACE codes (Rev. 2) as high- and medium-tech manufacturing (Table 2). The columns d and e of Table 1 provide the corresponding figures, and in column f, the values of $\Delta T$ for this subset (in column e) are compared with those in column c for all sectors. Columns c (for the total set) and e (for high- and medium-tech manufacturing) are highly correlated: Pearson $r = .962$ [$p < .01$]; Spearman's $\rho = .984$ [$p < .01$]; $N = 31$. However, the relative contribution of high- and medium-tech to the nation provides only 21.2% of the synergy, while 33.3% of the firms were classified as such.

The contributions are most pronounced in the regions of Beijing (27.9%), Shanghai (29.9%), and Tianjin (30.8%). The latter is a province between Beijing and the coast. In the next section, we shall see that these three provinces are also considered as municipalities with Chongqing as a fourth one. However, the knowledge base of this latter province is not so strongly enhanced given this focus on high- and medium-tech manufacturing.

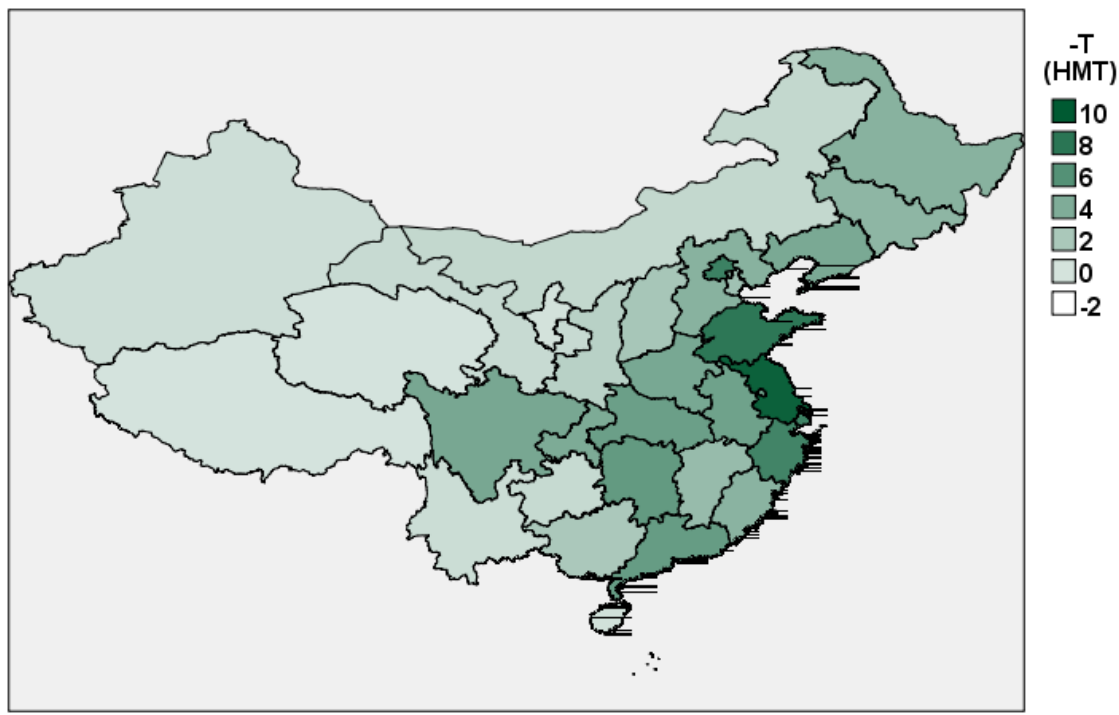

**Figure 4**: Synergies in high- and medium-tech manufacturing at the provincial level of China (years 2008-2010; *N* = 128,374).

*3.2.1 Knowledge-intensive services*

We found an uncoupling of the *knowledge-intensive services* from the regional economy in the Western-European countries studied previously because a knowledge-intensive service can be provided from any location near a railway station or airport; the geographical location ("rooting") is thus less relevant in the case of knowledge-intensive services. This effect is attenuated when R&D facilities are needed ("high-tech knowledge-intensive services") because these activities may require laboratories that are grounded. Since our data is thin in this domain (column g), we did not pursue a further decomposition within the category of knowledge-intensive services.

Column h shows that Chongqing is strongest in terms of this uncoupling effect with Beijing at the second place. The synergy indicator in these instances is positive which means that one adds to the uncertainty with a localized focus on these provinces. However, this is not the case for all provinces. Shanghai, for example, does not seem to play a role from this perspective (although the data is extremely poor; $N = 17$). Guandong which is represented with 346 of these services, however, does not perform any better on this indicator ($\Delta T = 0.01$ mbit).

In summary, the decomposition in terms of sectors most relevant to the knowledge-based economy did not show the strongly enhanced role of high- and medium-tech manufacturing that could be expected on the basis of previous studies (for European nations), but the uncoupling effect of knowledge-intensive services was confirmed for administrative centers such as Chongqing and Beijing. A further decomposition in terms of high-tech might be interesting, but we were hesitant to pursue this further given the limitations in this data.

*3.3 The second administrative level*

Figure 5 shows the pronounced contribution to the synergy at the second administrative level of four units (Beijing, Shanghai, Tianjin, and Chongqing) that have the special status of "municipalities" directly managed by the central government of China. Table 3 provides the numerical breakdown amd shows that the data is also mainly collected from these municipalities and the prefecture of Dezhou (neighboring to Tianjin, but in the province of Shandong) that follows at a next-lower level. However, Cangzhou—south of Tianjin—is larger in terms of the number of firms, but indicated as much less synergetic in terms of the three dimensions studied here.

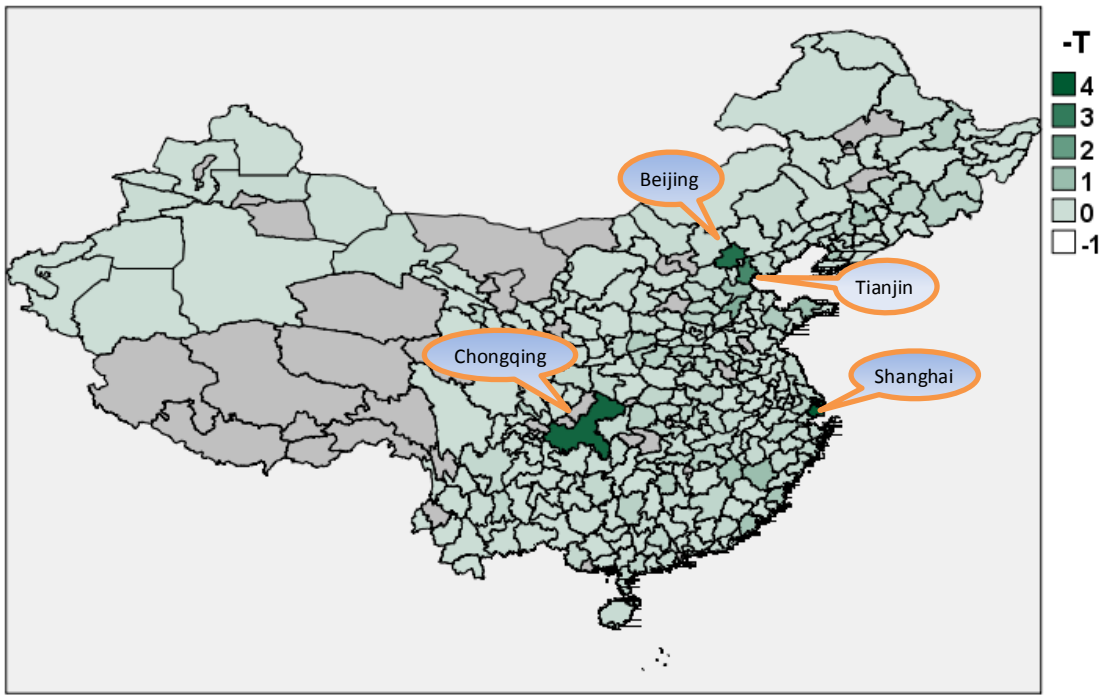

**Figure 5**: The distribution of 339 second-level administrative units in the PRC compared in terms of their contribution to the synergy among technology, geography, and organization.

|  | ΔT in mbits | n of firms |
|---|---|---|
| **Shanghai** | -3.91 | 12,742 |
| **Chongqing** | -3.65 | 13,488 |
| **Beijing** | -3.32 | 4,394 |
| **Tianjin** | -2.60 | 6,316 |
| Dezhou | -1.46 | 6,630 |
| Nanping | -0.96 | 1,823 |
| Yantai | -0.90 | 3,127 |
| Cangzhou | -0.82 | 7,628 |
| Zhangzhou | -0.75 | 4,830 |
| Fuzhou | -0.72 | 2,894 |
| Tieling | -0.67 | 2,349 |
| Weifang | -0.58 | 4,427 |
| Yuncheng | -0.51 | 735 |
| Zhengzhou | -0.51 | 1,900 |
| Hengyang | -0.47 | 998 |
| Deyang | -0.44 | 913 |
| Luohe | -0.44 | 2,994 |
| Yichun | -0.43 | 765 |
| Luoyang | -0.41 | 1,675 |
| Yanbian Korean | -0.38 | 227 |
| (…) | | |
| Σ | -40.84 | 310,974 |
| China (adm_2) | -183.42 | |
| $H_0$ | -142.58 | |

**Table 3**: Administrative units at the second level sorted in terms of their contribution (ΔT) to the synergy among technology, geography, and organization.

Because we can evaluate only 87.3% of the data at this level, the reduction of uncertainty at the national level is -183.42 as against -196.48 mbits in Table 1 (93.4%). Table 3 shows that only 22.3% of this synergy (40.84 mbits) is found at the second level of the administration. The provinces (at level 1) are thus the relevant units for studying the synergy in the Chinese economy, but the role of the national level is considerable. The country is far more centralized than, for example, Norway; but the distribution is less skewed towards the metropolitan centers than in Sweden.

**Discussion**

As noted, the major point for discussion is the data collection by the Bureau van Dijk that fills the *Orbis* database on the basis of information provided by more than one hundred information suppliers and by its own research. The methods of data-collection are discretionary since the information is controlled by the companies in question. This data, therefore, provides an incomplete sample. However, the selection of journals for example by Thomson-Reuters and Elsevier for their databases (Web of Science and Scopus, respectively) is also not public. Thus, such a state of affairs is more common when using commercial databases for scientometric research.

The main remaining question concerns the extent to which one can expect biases in the data-collection to influence the results. We noted that small-sized enterprises are under-represented in this data, and that self-employed entrepreneurs seem not to be included at all. Consequently, startups are presumably not included. This may bias the results against university-based entrepreneurship that may not be distributed equally across the country. However, the industrialized provinces of China all have a considerable number of universities with potentially industrial activities of graduates. This effect, however, may disfavor the largest metropolitan areas such as Beijing and Shanghai. (Hong Kong is not included in this data.)

The synergy indicator is not an output but a structural indicator, although—as noted—the number of firms is also registered in Eq. 2 as the units of analysis. Most governmental (OECD) statistics are (linear) output indicators (e.g., Schaaper, 2009). For example, we found a report entitled "High-tech statistics China 2012" on the website of the National Bureau of Statistics[4] which provides a table (1-7) with "Gross Industrial Output value of high-tech industries by region" (in RMB ¥100 million). Not surprisingly, this indicator correlates significantly with our indicator using the first-administrative level of 31 provinces; Spearman's $\rho = 0.867$ ($p < .01$). All distributions over the Chinese provinces are skewed and tend therefore to be correlated.

---

[4] Available at http://www.sts.org.cn/sjkl/gjscy/data2012/data12.pdf .

If we focus within the data only on high- and medium-tech, Spearman's ρ further increases only marginally to .879. However, the regions indicated as most productive in terms of output are different from the ones signaled by our methods, namely Jangsu and Guangdong. These latter regions are among those with the largest numbers of firms in our set: 62,805 and 44,692 firms, respectively. Zhejiang (50,699 firms in our data), however, is categorized with Beijing, Shanghai, and Shandong in the second group by this report, and with Chongqing among the third category.

As against these rankings in terms of size, our indicator is a measure of synergy or resonance among the distributions in three (or more) dimensions. The decomposition in terms of sectors and levels taught us that the four municipalities carry a specific function in the knowledge-based economy that was not anticipated. High- and medium-tech is more concentrated in and around the metropolitan areas of Beijing and Shanghai, but setting this filter does not lead to a higher synergy across the provinces (as we had expected on the basis of previous studies). The roles of Chongqing and Beijing as centers of knowledge-intensive administration are notable. In other words, the design allows us to fine-tune in terms of the three relevant dimensions (technology, geography, and size) as different projections of the three-dimensional data. The quality of this data, however, remains beyond our control.

**Conclusion**

Most synergy is generated in the knowledge base of the Chinese economy at the provincial level (administrative level 1), but the national level adds a substantial contribution (>18%). The level of the prefectures seems much less relevant except in the case of Shanghai, Beijing, Chongqing, and Tianjin as "municipalities" that are defined also at the (first) level of the provinces. In general, the synergy is generated above the city level; that is, at the level of regions.

A focus on high- and medium-tech manufacturing shows that the differences in terms of this selection can be between 20 and 30 percent for provinces. Tianjin joins Beijing and Shanghai as the "winners" when taking this perspective, whereas large industrial regions (such as Guandong) may be less profiled in terms of high- and medium-tech firms. Decomposition in terms of knowledge-intensive services is based on relatively small sets, but BvD claims that services are also included in *Orbis* and thus this data suggests that the Chinese economy is more manufacturing than service-oriented. The knowledge-intensive services are geographically located in the administrative centers (and perhaps associated to the government; cf. Perevodchikov *et al.*, 2013).

**Acknowledgement**
We thank Inga Ivanova, Fred Y. Ye, and two anonymous referees for comments on a previous version of this manuscript. The study was supported by the National Natural Science Foundation of China (NSFC) with grant number 71073153.